%% file: AD_Paper_v2.tex
\documentclass[sigconf]{acmart}

\usepackage{booktabs} 
\usepackage[T1]{fontenc}
\usepackage[latin9]{inputenc}
\usepackage{array}
\usepackage{rotating}
\usepackage{float}
\usepackage{multirow}
\usepackage{amsmath}
\usepackage{amssymb}
\usepackage{graphicx}
\usepackage{setspace}
\usepackage[linesnumbered,ruled]{algorithm2e}

\providecommand{\tabularnewline}{\\}
\copyrightyear{2018}
\acmYear{2018}
\setcopyright{acmlicensed}
\acmConference[JCDL'18]{The 16th ACM/IEEE Joint Conference on Digital Libraries}{June 3--7, 2018}{Fort Worth, TX, USA}
\acmBooktitle{JCDL '18: The 18th ACM/IEEE Joint Conference on Digital Libraries, June 3--7, 2018, Fort Worth, TX, USA}
\acmPrice{15.00}
 \acmDOI{10.1145/3197026.3197036}
\acmISBN{978-1-4503-5178-2/18/06}

\begin{document}

\fancyhead{}

\title{Effective Unsupervised Author Disambiguation\\with Relative Frequencies}

\author{Tobias Backes}
\affiliation{  \institution{GESIS - Leibniz-Institute for the Social Sciences } }
\email{tobias.backes@gesis.org}

\input{AD_abstract}
\input{AD_concepts}


\maketitle

\input{AD_body_v2}

\section*{Acknowledgments} This work was supported by the EU's Horizon 2020 programme under grant agreement H2020-693092, the \hyperlink{http://moving-project.eu}{MOVING} project.

\bibliographystyle{ACM-Reference-Format}
\bibliography{d3.1/GESIS} 

\input{AD_appendix}

\end{document}

%% file: AD_abstract.tex
\begin{abstract}
This work addresses the problem of author name homonymy in the Web of Science.
Aiming for an efficient, simple and straightforward solution, we introduce a novel probabilistic similarity measure for author name disambiguation based on feature overlap.
Using the researcher-ID available for a subset of the Web of Science, we evaluate the application 
of this measure in the context of agglomeratively clustering author mentions.
We focus on a concise evaluation that shows clearly for which problem setups and at which time during the clustering process our approach works best.
In contrast to most other works in this field, we are skeptical towards the performance of author name disambiguation methods in general and compare our approach to the trivial single-cluster baseline.
Our results are presented separately for each correct clustering size as we can explain that, when treating all cases together, the trivial baseline and more sophisticated approaches are hardly distinguishable in terms of evaluation results.
Our model shows state-of-the-art performance for all correct clustering sizes without any discriminative training and with tuning only one convergence parameter.
\end{abstract}

%% file: AD_concepts.tex
\ccsdesc[500]{Information systems~Entity resolution}

%% file: AD_body_v2.tex
\section{Introduction}

Documents have authors. This information is almost always available
on a document and in the document's metadata. However, it is crucial
to distinguish an author name mentioned on a specific document from
the author itself. Usually, the author is \emph{referred to} by a
string of characters that is given with the document. This concept
introduces two types of ambiguity:
\begin{enumerate}
\item \emph{Name synonymy}: One author is referred to by different 
strings, perhaps due to misspelling, language-specificity,
different conventions for name specification, etc.
\item \emph{Name homonymy}: One string refers to different authors.
With the the collection size the chance increases that two authors in the collection have the same name.
\end{enumerate}
In general, both problems have to be addressed simultaneously, i.e.
in a constrained clustering setup. This allows to establish that for
example \emph{DOE, J} can be \emph{DOE, JW} or \emph{DOE, JH} but
not both. In this contribution, we focus on name homonymy and simplify
the problem by considering that all initials are always given (assuming
that \emph{DOE, J}, \emph{DOE, JW} and \emph{DOE, JH} are different
persons). In that case, \emph{Author(ship/name) disambiguation} decides
for a set of \emph{author mentions} with the same name, which of them
belong to the same author and which do not. This is a clustering problem
over author mentions. Each cluster is considered an author. More formally:
\begin{itemize}
\item For each collection, there is a set $\mathfrak{N}$ of \emph{names}
$name\in\mathfrak{N}$
\item For each name $name$, there is a set $\mathfrak{C}$ (also referred
to as a \emph{clustering}) of authors $C\in\mathfrak{C}$ (also referred
to as a \emph{cluster}) and a set of mentions $x\in X$
\item Each author $C\subseteq X$ is a set of \emph{mentions} $x\in C$
\item For each mention $x$, there is a bag of \emph{features} $f\in F(x)$,
each with a frequency $\#(f,x)$ of occurrence with $x$
\end{itemize}

\begin{figure}[b]
\begin{centering}
\includegraphics[height=1.2cm]{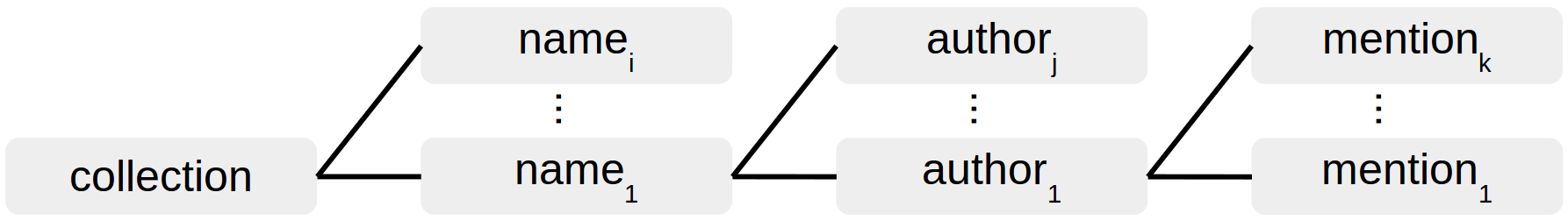}
\par\end{centering}
\caption{Author disambiguation problem structure}
\end{figure}

If a name is not disambiguated, we have no information about the belonging
of single mentions. This state can either be expressed by putting
all mentions in the same cluster $C=X$ such that $\mathfrak{C}=\{\{x\mid x\in X\}\}$,
or by assigning each mention $x$ its own cluster $C=\{x\}$, such
that $\mathfrak{C}=\{\{x\}\mid x\in X\}$. The task of author disambiguation
is then to suggest a \emph{system} clustering $\mathfrak{C}_{sys}$
that is as close to the correct clustering $\mathfrak{C}_{cor}$ as
possible. In the training/tuning and evaluation case, we have both
$\mathfrak{C}_{sys}$ and $\mathfrak{C}_{cor}$ present and optimize
some evaluation score $eval(\mathfrak{C}_{sys},\mathfrak{C}_{cor})$.
We can safely assume that this score measures the similarity between
the system- and the correct clustering. In practice, we disambiguate one name at a time. This is referred
to as \emph{blocking} \cite{torvik2009author}, where each name defines one \emph{block}, that is a separate problem set.

Many different approaches have been proposed to solve the problem of author 
disambiguation. Although it is often not indicated as clearly, most of these approaches can 
be viewed within the framework and terminology described above. Despite the large number 
of proposed methods, we find that a straightforward, simple and sound, easy-to-implement 
-- yet well performing -- baseline is still lacking. Existing solutions are either complicated to 
understand and implement, require training steps or are based on a number of rather arbitrary 
decisions. Normally, performance is not compared against the most primitive conceivable baseline \cite{milojevic2013accuracy}. Therefore, in this work, we elaborate on the following research questions:
\begin{description}
\item [{RQ1}] Can we deploy well-behaved, straightforward probabilities to establish
a conceptionally simple, fast and reliable method for author name
disambiguation that shows state-of-the-art performance?
\item [{RQ2}] How difficult is the problem of author name disambiguation
in general and separately for different problem sizes? How good are
the most primitive baselines and how does our approach compare to
them?
\end{description}
We structure the rest of the paper as follows. In section 2, we review
the related work. In section 3, we describe the probabilistic model
that we designed as a proof-of-concept for RQ1. In section 4, we explain
the evaluation setup, including simple baselines and the
separation of different problem sizes. This allows us to test
our approach regarding RQ1 and gather insights towards the more
general RQ2. We summarize the most prominent findings regarding these
two research questions in section 5.

\section{Related work}

In this section, we give a brief overview over the most cited literature
on author disambiguation that relates to our approach:

\paragraph{The Problem of Author Disambiguation}

Ferreira et al. \cite{ferreira2012brief} give an overview of author
disambiguation, distinguish author grouping (\emph{synonymy}) vs.
author assignment (\emph{homonymy}) and different types of features
for the latter. Smalheiser and Torvik \cite{smalheiser2009author}
go into more detail, but to some extend also concerning their own project;
Harzing \cite{harzing2015health} analyses the top 1\% cited academics
from Thompson Reuters Essential Science Indicators and contributes
some details of the ambiguity problem in different languages. She
shows that the extent of ambiguity has a direct influence on the scientific
performance indicators measuring a scientist's academic output. Strotmann
and Zhao \cite{strotmann2012author} investigate the application of
author disambiguation to citation networks. They show
that  one of the reasons for researchers being top-ranked is in fact 
a lack of author name disambiguation. Kramer, Momeni and Mayr \cite{2017arXiv170301319K}
give an overview of the quantity of author gold annotation in the
Web of Science. They conclude that the Web of Science researcher-ID
used in this paper is a good source of author identity information,
in contrast to other sources. Milojevic \cite{milojevic2013accuracy}
analyzes the accuracy of name blocks built by considering
all available initials of an author (and his last name). On the small
sample used, this accuracy introduces a very high baseline.

\paragraph{Probabilistic Approaches, Topic Models}

Han et al. \cite{han2005hierarchical} present a probabilistic Naive Bayes mixture model to disambiguate a small set of highly ambiguous author names. They compare to K-means and find that their model performs significantly better. However, they use a very long product, which can lead to extreme swings and the EM algorithm, a computationally and conceptionally rather complex method. 
Tang et al. \cite{tang2012unified} suggest a general probabilistic model based on Markov Random Fields to exploit a variable set of features for author disambiguation. They evaluate their approach on a small set of highly ambiguous names and compare the results to more basic clustering techniques, claiming significant improvements. With \textasciitilde{}89\% F1, their results are indeed impressive, but considerable effort is required to understand and implement the method.
Torvik et al. \cite{torvik2005probabilistic} present a probabilistic model that compares pairs of authors based on various features such as terms, affiliations or venues. Torvik and Smalheiser \cite{torvik2009author} apply an improved version of this previous approach to the Medline data set and reach  \textasciitilde{}95\% F1 for the one name reported. Their work is one of the most cited state-of-the-art methods in this field. Generally speaking, their experience allows the group to hard-code a lot of knowledge into their methods, but not all decisions can be easily comprehended by others. Although they use some probabilities in their formulas, it seems the final similarities are not normalized.
Song et al. \cite{song2007efficient} present a probabilistic graphical topic model for hierarchical clustering of author names and compare their approach to other standard clustering approaches. They find their model performs better than for example DBSCAN. Although the good performance of  \textasciitilde{}90\% F1 seems to justify the additional complexity, like Han et al., they introduce long products and inference steps on a 'detour' via topic assignments. 

\paragraph{The Web of Science, Training Techniques}

Like us, Gurney et al. \cite{gurney2012author} work on the Web of Science and use a very similar set of features. They deploy another group's method for clustering and the Taminoto coefficient for similarity.  They report F1$>$90\% for a number of names. Their approach is the one most similar to our proposal, but neither does it use feature-specificity nor return probabilities.
Levin et al. \cite{levin2012citation} present a semi-supervised approach to author disambiguation on the Web of Science data set. The evaluation of their approach shows solid performance (but under 90\% F1) and proves that the method scales to very large collections. As they do pairwise classification of author mentions, they make the problem harder than it needs to be. Overall, their work is very detailed but also quite difficult to reproduce.
Ferreira et al. \cite{ferreira2010effective} present a semi-supervised classifier evaluated on DBLP and BDBComp data sets. Their model outperformed the other methods compared. Depending on supervision and rules, their model introduces a number of (dataset-specific) parameters. Recently, this group has updated their method to allow for incremental disambiguation by comparing not author mentions but mentions to clusters of mentions. \cite{santana2017incremental} This is very practical but more difficult.
In another paper focusing on training aspects, Culotta et al. \cite{culotta2007author} present a special similarity function and combine error-driven sampling with learning-to-rank. They also give an overview of other training approaches. They find that their approach can be beneficial in terms of performance. Similar to Ferreira  et al., this approach requires training and uses specific similarity measures for different feature types.

\section{Method description}

We intent to define a simple, yet effective probabilistic method for author disambiguation.
Using the blocking paradigm, our method disambiguates one name at
a time. It clusters all mentions of that name based on features extracted from the collection.
We note that for each mention $x$, there is exactly one document $d(x)$
in which this mention appears. However, for each document, there can
be multiple mentions that appear on it.

\subsection{Features}

Features $F(x)$ assigned to a mention $x$ can be extracted (1) from
$d(x)$ in general or (2) specifically for $x$ on $d$. In the first case, the features $F(x)$ 
are the same as the features $F(\dot{x})$ of a mention $\dot{x}$ that appears
on $d$ as well. In the following, we will distinguish different \emph{feature-types},
some of which are \emph{unspecific} for $x$ and some
of which are \emph{specific}. We train and test
our approach on the Web of Science, as it is the major source of scientific
documents annotated with actual author IDs (\emph{researcher-ID}).
The following information is used:
\begin{enumerate}
\item \emph{Terms} $F_{term}(x)$: All words considered relevant in the document and their frequency of occurrence in $d(x)$
\item  \emph{Affiliations} $F_{aff}(x)$): All affiliations given for $x$ on $d(x)$ -- usually just one
\item \emph{Categories} $F_{cat}(x)$: All categories assigned to $d(x)$, where we consider categories to be relatively general terms picked from a relatively small vocabulary / thesaurus. The frequency of one category for a document $d$ is usually $1$
\item \emph{Keywords} $F_{key}(x)$: All keywords assigned to $d(x)$, where we consider keywords to be relatively specific terms that are picked from a relatively large vocabulary or only with respect to the current document. The frequency of one keyword for a document $d$ is usually $1$
\item\emph{Coauthornames}  $F_{co}(x)$: All names of the coauthors of $x$ on $d(x)$. Unless more than one author of $d$ have the same name, the frequencies are $1$
\item \emph{Refauthornames} $F_{ref}(x)$: All author names from documents $d'$ in the collection such that $d$ references $d'$. Frequencies larger than $1$ will occur often, for example if multiple documents by the same author are referenced
\item \emph{Emails} $F_{email}(x)$: All email addresses given for $x$ on $d(x)$ -- usually just a single email address with a frequency of $1$
\item \emph{Years} $F_{year}(x)$: A bag of years $f$ given for $d(x)$. Like  Levin et al. \cite{levin2012citation}, we model $\#(f,x)$ as a Gaussian with   the publication year of $x$ as mean. This models temporal proximity.
\end{enumerate}
While there are many details related to the question of which and
how features are extracted and normalized, the focus of our research
was not to investigate specific features but to develop a method that
can provide satisfying results independent of the exact set of features
and feature-types.

\subsection{Agglomerative clustering}

\input{AD_algorithm}

We apply a method of agglomerative clustering as we consider it the most straightforward approach.
This means that we start with the initial state where each mention $x$ is in its own
cluster $C=\{x\}$. Then, pairs $(C,\dot{C})$ of clusters are merged.
If no stopping criterion is applied, this will ultimately result in
a state where all mentions are in the same cluster $C=X$. For this
reason, we need to compute the score $score(C,\dot{C})$ of a pair
$(C,\dot{C})$ of clusters to be merged. Furthermore, we deploy a
\emph{quality limit} $l$, that tells us whether the score can be
considered good or not. In our approach, $score(C,\dot{C})$ is not
dependent on the score of any other pair of clusters. Neither is the
quality limit. This means that in each iteration of the clustering
process, we merge all pairs $(C,\dot{C})$, such that (1) $\neg\exists\ddot{C}\in\mathfrak{C}:score(C,\ddot{C})>score(C,\dot{C})\wedge\neg\exists\dddot{C}\in\mathfrak{C}:score(\dddot{C},\dot{C})>score(C,\dot{C})$
and at the same time, (2) $score(C,\dot{C})>l$. In other words, we
evaluate all disjoint pairs $(C,\dot{C})\in\mathfrak{C}\times\mathfrak{C}$;
for each of these pairs, we check whether (1) and (2) hold true. If
yes, the pair is saved for merging. See algorithm \ref{alg:1} for
a more formal description. At the end of each iteration, all saved
pairs are merged. A new system clustering is obtained and the next
iteration begins. This process converges if no pairs are saved for
merging. For evaluation purposes, we can continue to merge with moves
that are below the limit, but we will elaborate on this in the experiments
section.

\subsection{Probabilistic similarity}

The main contribution of our approach is the similarity used to define
$score(C,\dot{C})$. In the following, we will define and explain
the probability that inspires the score. We say that the score of
a pair of clusters can be seen as their \emph{joint probability} $p(C,\dot{C})=p(\dot{C},C)=p(C|\dot{C})\cdot p(\dot{C})$.
Remember $\#(f,x)$ denotes the frequency of $f$ in the set $F(x)$
of all features in $x$ ($F=\bigcup_{x\in X}F(x)$). Consider $\#(f,x)=0$
if $f\notin F(x)$. We define:\allowdisplaybreaks
\begin{align*}
p(C|\dot{C}) & =\sum_{(x,\dot{x})\in C\times\dot{C}}p(x|\dot{x})\cdot\frac{\#(\dot{x})}{\#(\dot{C})} & p(C) & =\sum_{x\in C}p(x)\\
p(x|\dot{x}) & =\sum_{f\in F}\frac{\#(f,x)\cdot\#(f,\dot{x})}{\#(f)\cdot\#(\dot{x})} & p(x) & =\frac{\#(x)}{\#(\centerdot)}\\
\#(C) & =\sum_{x\in C}\#(x) & \#(x) & =\sum_{f\in F}\#(f,x)\\
\#(f) & =\sum_{x\in X}\#(f,x) & \#(\centerdot) & =\sum_{x\in X}\#(x)
\end{align*}
To prevent division by zero, we apply $add\epsilon$ smoothing. This
modifies $p(C|\dot{C})$, $p(\dot{x})$ and $p(x|\dot{x})$:\allowdisplaybreaks
\begin{align*}
p(C|\dot{C}) & =\sum_{(x,\dot{x})}p(x|\dot{x})\cdot\frac{\#(\dot{x})+\epsilon}{\#(\dot{C})+|C|\cdot\epsilon} & p(x) & =\frac{\#(x)+\epsilon}{\#(\centerdot)+|X|\cdot\epsilon}
\end{align*}
\begin{align*}
p(x|\dot{x}) & =\frac{1}{\#(\dot{x})+\epsilon}\cdot\left(\left(\sum_{f\in F}\frac{\#(f,x)\cdot\#(f,\dot{x})}{\#(f)}\right)+\frac{\epsilon}{|X|}\right)
\end{align*}

\subsection{Variations}

So far, we have considered the clustering to be fully enclosed in
the current name block. While still treating each name as a separate
clustering problem, we can say that $X$ is not only the set of mentions
for the current name, but for the entire collection. This increases
$\#(f)$, $\#(\centerdot)$ and obviously $|X|$. We find that the
performance is considerably better if this approach is taken, which
can be explained with less sparsity for $\#(f)$. However, this poses
the question, what the 'collection' is that we take these counts from.
Basically, we are simply looking for a realistic distribution of the
frequency of features $f$ independent of the mention they occur with.
In an application scenario, this distribution can be taken from the
data to be clustered itself. As $\#(f)$, $\#(\centerdot)$ and $|X|$
are available even for the data we cluster, in our evaluation scenario,
we obtain them from the union of the training and testing portion.

In our preferred variant, we only use $p(C|\dot{C})$ in $score(C,\dot{C})$.
This conditional probability does in fact perform much better than
the joint probability. Intuitively, $p(C)$ favours large clusters
for merging, which does not make sense as it introduces a tendency
that reinforces itself. It does not matter that $p(C|\dot{C})\neq p(\dot{C}|C)$
as merging is symmetric and the clustering procedure we described
above will simply use $max(p(C|\dot{C}),p(\dot{C}|C))$ unless there
is a third cluster that matches even better (or the limit is not met).
In this case, $p(\dot{x})$ is never applied, so that we do not need
$\#(\centerdot)$. Collecting $\#(f)$ over the entire collection,
we then only need the number $N$ of mentions in the collection. 
Therefore, $X$ still denotes the set of mentions for one
single name block. Furthermore, we tested a variant where we replace
the sum of products with a maximum of products:
\[
\tilde{p}(C|\dot{C})=\max_{(x,\dot{x})\in C\times\dot{C}}p(x|\dot{x})\cdot\frac{\#(\dot{x})}{\#(\dot{C})}
\]
Obviously, the value $\tilde{p}$ inspired by single-link clustering
is not a probability anymore. This variant performed slightly different
than the conditional probability.

\subsection{Feature-type weights}

All the probabilities shown above are obtained \emph{separately} for
each feature-type. Consider that each probability $p$ should actually
be denoted $p_{ftype}$ where $ftype$ is either \emph{term}, \emph{aff},
\emph{cat}, \emph{key}, \emph{co}, \emph{ref}, \emph{email} or \emph{year}.
For better readability, we drop the subscript where it is not necessary.
We perform a simple linear combination with feature-type weights $\lambda$
to obtain the final score:
\[
score(C,\dot{C})=\sum_{ftype}\lambda_{ftype}\cdot p_{ftype}(C|\dot{C})
\]
Ideally, we would like to avoid all training so as to keep our method as simple as possible. 
Still, to allow for comparison, weights $\lambda$ are trained on the training portion of our data.
We sample pairs $(x,C)$ and $(x,\dot{C})$ such that $x\in C\wedge x\notin\dot{C}\wedge|C|=|\dot{C}|\wedge C\cup\dot{C}\subseteq X$,
where $X$ is the set of mentions for a \emph{single} name. All possible
values for $|C|=|\dot{C}|$ are considered in order to create a more
or less realistic binary classification scenario, where we are asked
to assign $x$ to a correct cluster $C$ or an incorrect cluster $\dot{C}$.
The classifier (\emph{logistic regression} performed well) receives
probabilities $p_{ftype}(x|C)$ and $p_{ftype}(x|\dot{C})$ for each
$ftype$ together with the class 'correct' or 'incorrect'. It then
learns feature-type weights $\lambda_{ftype}$ in order to optimize
the classification outcome. While this is not the same scenario as
the one that the weights are finally applied in, we hope that nevertheless,
we gather some insight into the importance of single feature-types.

\subsection{Convergence}

Above, we have introduced a quality limit $l$ on the scores for moves
during clustering. In order to account for different problem sizes
(and corresponding smaller probabilities), we define $l$ as follows:
\[
l=\alpha+|X|\cdot\beta
\]
where $|X|$ is the number of mentions for the current name. Fortunately,
when they are normalized such that $\sum_{ftype}\lambda_{ftype}=1$,
$l$ is relatively independent of $\lambda$. Another pleasant finding
was that it is sufficient to tune one parameter depending on whether
we use the sum-of-products or the maximum-of-products variant. 
While we optimize both parameters, results were best if in the first case  $\beta=0$ and in the second $\alpha=0$.

\begin{figure}
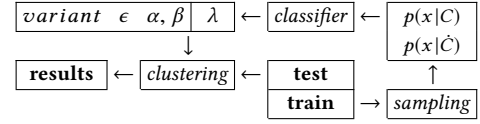

\noindent \begin{centering}
\setlength\tabcolsep{2pt}
\begin{tabular}{|ccc|c|c|c|c|c||c|}
\cline{1-4} \cline{6-6} \cline{8-9} 
\multirow{1}{*}{{\small{}$variant$}} & \multirow{1}{*}{{\small{}$\epsilon$}} & {\small{}$\alpha,\beta$} & \multirow{1}{*}{{\small{}$\lambda$}} & {\small{}$\leftarrow$} & \emph{\small{}classifier} & \multirow{1}{*}{{\small{}$\leftarrow$}} & \multicolumn{2}{c|}{{\footnotesize{}$p(x|C)$}}\tabularnewline
\cline{1-4} \cline{6-6} 
\multicolumn{1}{c}{} &  & \multicolumn{2}{c}{{\small{}$\downarrow$}} & \multicolumn{1}{c}{} & \multicolumn{1}{c}{} &  & \multicolumn{2}{c|}{{\footnotesize{}$p(x|\dot{C})$}}\tabularnewline
\cline{1-1} \cline{3-4} \cline{6-6} \cline{8-9} 
\multicolumn{1}{|c|}{\textbf{\small{}results}} & \multicolumn{1}{c|}{{\small{}$\leftarrow$}} & \multicolumn{2}{c|}{\emph{\small{}clustering}} & {\small{}$\leftarrow$} & \textbf{\small{}test} & \multicolumn{1}{c}{} & \multicolumn{2}{c}{{\small{}$\uparrow$}}\tabularnewline
\cline{1-1} \cline{3-4} \cline{6-6} \cline{8-9} 
\multicolumn{1}{c}{} &  & \multicolumn{1}{c}{} & \multicolumn{1}{c}{} &  & \textbf{\small{}train} & {\small{}$\rightarrow$} & \multicolumn{2}{c|}{\emph{\small{}sampling}}\tabularnewline
\cline{6-6} \cline{8-9} 
\end{tabular}
\par\end{centering}
\caption{Setup for clustering, evaluation and training $\lambda$  \label{fig:overview}}
\end{figure}

\subsection{Implementational Details}

Our approach as presented in previous sections can be implemented
in an efficient and conceptionally simple way by means of matrix multiplication.
In a first step, we calculate the matrix $\overline{p(x|\dot{x})}$ containing
values $p(x|\dot{x})$ for all pairs $(x,\dot{x})\in X\times X$. This is based on the $|X|\times(|F|+1)$
count matrix $\overline{\#(x,f)}$, the feature count vector $\overrightarrow{\#(f)}$, and the mention count vector $\overrightarrow{\#(x)}$. 
For smoothing, $\overline{\#(x,f)}$ is extended by one column $\langle\epsilon\ldots\epsilon\rangle^T$.
\[
\begin{array}{cc}
\overline{p(x|\dot{x})}=\overline{p(x|f)}\cdot\overline{p(f|\dot{x})} & \overline{p(x|f)}=\frac{\overline{\#(x,f)}}{\overrightarrow{\#(f)}^{T}}\\
\overline{p(f|\dot{x})}=\frac{\overline{\#(x,f)}}{\overrightarrow{\#(x)}} & \forall i:\overline{\#(x,f)}_{i,|F|+1}=\epsilon
\end{array}
\]
Here, $\cdot$ denotes matrix multiplication (\emph{matrix- or dot
product}). During each iteration, we calculate the $|\mathfrak{C}_{sys}|\times|\mathfrak{C}_{sys}|$
matrix $\overline{p(C|\dot{C})}$ from $\overline{p(x|\dot{x})}$ and the current
clustering as a $|X|\times|\mathfrak{C}_{sys}|$ matrix $\overline{\mathfrak{C}_{sys}}$:
\[
\begin{array}{cc}
\overline{p(C|\dot{C})}=\overline{\mathfrak{C}_{sys}}^{T}\cdot\left(\overline{p(x|\dot{x})}\cdot\overline{p(\dot{x}|C)}\right) & \overrightarrow{\#(C)} =\overrightarrow{\#(x)}^{T}\cdot\overline{\mathfrak{C}_{sys}}
\end{array}
\]
\[
\begin{array}{c}
\overline{p(\dot{x}|C)}=\left(\left(\overrightarrow{\#(x)}+\epsilon\right)\cdot\frac{1}{\overrightarrow{\#(C)}+|\overline{\mathfrak{C}_{sys}}|\epsilon}\right)\circ\overline{\mathfrak{C}_{sys}}
\end{array}
\]
Here, $\circ$ denotes component-wise multiplication (\emph{Hadamard
product}), ensuring $p(x|C)=0$ if $x\notin C$.
For the \emph{max} variant, all sums in the matrix products are replaced
by a maximum function.

\section{Experimental evaluation}

In the following we describe the experimental setup used to train,
tune and test our approach. This is also depicted in figure \ref{fig:overview}.

\subsection{Data}

We use the \emph{Web of Science} (\emph{WoS}) collection as a source
of metadata and annotated authorship information. In a preprocessing
step, we extract features for the feature-types mentioned in the previous
section. We normalize author names as LASTNAME, INITS, where I, N,
I, T, S are all the initials for each first name of the author mention.
Thereby, on the one hand, we make the problem harder as we drop the
full name information (increasing the problem size, i.e. \emph{John
Doe} $=$ \emph{Jack Doe}) and on the other we make it simpler as
we treat mentions of the same author where his first names are given
with varying completeness as separate problems (reducing the problem
size, i.e. \emph{John Doe} $\neq$ \emph{John W. Doe}). We extract
terms and their frequency from the title and abstract of the metadata.
Title terms are weighted three times higher than abstract terms. Some
stop words are omitted and all words are lower-cased. Furthermore some
basic lemmatization from the \emph{Natural Language Toolkit} (\emph{NLTK})
is applied. Affiliations are already normalized in the WoS. Categories
and keywords are taken as they are in the WoS. Co- and referenced author
names are normalized as described above. Emails are not normalized,
but we plan to lowercase them in the future. From the publication
date, we only pick the year.

After extracting features for the entire WoS with more than 100 million
documents, we create a database containing the features related to
each single mention with a researcher-ID. For each name, we use all
the mentions that are given a researcher-ID and we use all the authors
that contain at least one such mention. As stated earlier, we consider
each researcher-ID a distinct author. Names are ordered randomly and
separated into training and testing portions. We use 25\% names for
training.

\subsection{Test setup}

In contrast to work by most other groups, we are specifically interested
in evaluating the performance of our model in relation to the problem
size. If this has been done in the literature, it is usually regarding
the number $|X|$ of mentions with the same name (i.e. in Gurney et al. \cite{gurney2012author}). However, we sort
our name blocks by the correct number of clusters that would need
to be detected in order to achieve perfect results (the size $|\mathfrak{C}_{cor}|$
of the cluster\emph{ing}). We compare a maximum of $1000$ names for
all $|\mathfrak{C}_{cor}|\in\{1..10\}$. Of this selection, $25\%$
are used for training. Clustering sizes are distributed according to
Zipf's law. For the sizes $1$ to $4$, more than $1000$ names are
available. See table \ref{tab:The-number-of} for exact number of
names for each clustering size. We evaluate our approach for each
size separately. Doing so, we are basically balancing our data in
an artificial way. We find this is necessary, as the Zipf distribution
of $|\mathfrak{C}|$ leads to a (usually unobserved) preference of
models that create very few clusters. Furthermore, in order to view
the actual behavior of our method, we carefully monitor the development
of precision, recall and F1 measure with each iteration of the clustering
process. We also monitor how the clustering would have continued if
there were no limit $l$ on the score of possible merges. So for each
iteration in the clustering process, we record the following information:
\begin{enumerate}
\item \textbf{Precision} of current system clustering
\item \textbf{Recall} of current system clustering
\item Current \textbf{number of clusters} in system clustering
\item Whether the current iteration is before \textbf{convergence}
\end{enumerate}
During the clustering process, we continuously apply the quality limit
$l$ to merges. Once an iteration is reached where no possible merge
exceeds the limit, the following iterations are only hypothetical
and all possible merges are applied if and only if no merge exceeds
the limit. Note that this means that even during hypothetical iterations,
there might be a limit-based selection of merges. As there
is always at least one possible merge, all clusters are finally merged
into one and we can evaluate the development of precision and recall
as well as the point of convergence. This is particularly interesting
if one has certain preferences towards precision or recall and would
like to find a good stopping point for the clustering process. Figure \ref{fig:trained_details} 
shows how we plot the recorded information.

\begin{table}
\begin{centering}
\tabcolsep=1.9pt\renewcommand{\arraystretch}{1.1}%
\begin{tabular}{r|ccccccccccc}
\textbf{\small{}$|\mathfrak{C}|$} & {\small{}$1$} & {\small{}$2$} & {\small{}$3$} & {\small{}$4$} & {\small{}$5$} & {\small{}$6$} & {\small{}$7$} & {\small{}$8$} & {\small{}$9$} & {\small{}$10$} & {\small{}$\sum$}\tabularnewline
\hline 
\textbf{\small{}train} & \emph{\small{}255} & \emph{\small{}250} & \emph{\small{}250} & \emph{\small{}250} & \emph{\small{}215} & \emph{\small{}139} & \emph{\small{}80} & \emph{\small{}65} & \emph{\small{}43} & \emph{\small{}35} & \emph{\small{}1582}\tabularnewline
\textbf{\small{}test} & \emph{\small{}767} & \emph{\small{}751} & \emph{\small{}750} & \emph{\small{}750} & \emph{\small{}645} & \emph{\small{}418} & \emph{\small{}242} & \emph{\small{}195} & \emph{\small{}131} & \emph{\small{}106} & \emph{\small{}4749}\tabularnewline
\textbf{\small{}train+test} & \emph{\small{}1022} & \emph{\small{}1001} & \emph{\small{}1000} & \emph{\small{}1000} & \emph{\small{}860} & \emph{\small{}557} & \emph{\small{}322} & \emph{\small{}260} & \emph{\small{}174} & \emph{\small{}141} & \emph{\small{}6331}\tabularnewline
\textbf{\small{}all} & \emph{\small{}229653} & \emph{\small{}14108} & \emph{\small{}3630} & \emph{\small{}1657} & \emph{\small{}860} & \emph{\small{}557} & \emph{\small{}322} & \emph{\small{}260} & \emph{\small{}174} & \emph{\small{}141} & \emph{\small{}251362}\tabularnewline
\textbf{\small{}\% used} & \emph{\small{}0.45} & \emph{\small{}7.10} & \emph{\small{}27.55} & \emph{\small{}60.35} & \emph{\small{}100} & \emph{\small{}100} & \emph{\small{}100} & \emph{\small{}100} & \emph{\small{}100} & \emph{\small{}100} & \emph{\small{}2.52}\tabularnewline
\end{tabular}
\par\end{centering}

\caption{The number of names found or used with a correct clustering size $|\mathfrak{C}_{corr}|\in\{1..10\}$
in the WoS data \label{tab:The-number-of}}
\end{table}

\subsection{Evaluation measures}

In order to evaluate the performance of our approach, we use two popular
evaluation measures for clustering: (1) \emph{pairwise F1} (\emph{pairF1})
and (2) \emph{bCube}. Both measures define \emph{precision} (\emph{P})
and \emph{recall} (\emph{R}) when comparing two clusterings $\mathfrak{C}_{sys}$
and $\mathfrak{C}_{cor}$. F1 is defined as usual as $2\cdot\frac{P\cdot R}{P+R}$.
Except for one aspect, we use the definition by Levin et al. (2012)
\cite{levin2012citation}:\allowdisplaybreaks
\begin{align*}
pairs(\mathfrak{C}) & =\bigcup_{C\in\mathfrak{C}}\left\{ \{x,\dot{x}\}\mid x,\dot{x}\in C\wedge x\neq\dot{x}\right\} \\
P_{pairF1} & =\frac{pairs(\mathfrak{C}_{cor})\cap pairs(\mathfrak{C}_{sys})}{pairs(\mathfrak{C}_{sys})}\\
R_{pairF1} & =\frac{pairs(\mathfrak{C}_{cor})\cap pairs(\mathfrak{C}_{sys})}{pairs(\mathfrak{C}_{cor})}\\
C_{sys}(x) & =C\in\mathfrak{C}_{sys}:x\in C\\
P_{bCube} & =\frac{1}{|X|}\cdot\sum_{x\in X}\frac{|C_{sys}(x)\cap C_{cor}(x)|}{|C_{sys}(x)|}\\
R_{bCube} & =\frac{1}{|X|}\cdot\sum_{x\in X}\frac{|C_{sys}(x)\cap C_{cor}(x)|}{|C_{cor}(x)|}
\end{align*}

\noindent We note that the above shown $P_{pairF1}$ and $R_{pairF1}$
are not defined if in the first case $\mathfrak{C}_{sys}$ and in
the second $\mathfrak{C}_{cor}$ are $\left\{ \{\{x\}\mid x\in X\}\right\} $.
Therefore we modify $pairs(\mathfrak{C})=\bigcup_{C\in\mathfrak{C}}\left\{ \{x,\dot{x}\}\mid x,\dot{x}\in C\right\} $.
This increases the values for pairF1 slightly.

One important question with regard to these evaluation measures is
on which subset of the problem they are applied. It is understood
from the above formula, that there is a distinct precision and recall
value for each clustering problem, that is for each name. However,
one could also consider the pairs to be taken over the entire test
data:
\[
pairs_{cor}(\mathfrak{N})=\bigcup_{name\in\mathfrak{N}}\bigcup_{C\in\mathfrak{C}_{cor}^{name}}\left\{ \{x,\dot{x}\}\mid x,\dot{x}\in C\wedge x\neq\dot{x}\right\} 
\]
In that case one would calculate one value of precision and recall
over all correct and incorrect pairs in the test data. From these
two values, F1 could be calculated. If we calculate precision and
recall for each name separately, we have to average over the results
that we get for each single name. This weights each name equally (independent
of the number $|X|$ of mentions with that name). We can then calculate
F1 from the average precision and average recall over all names. We
use this approach to obtain a final score for each correct number
$|\mathfrak{C}|$ of clusters. We do not report a final score over
all $|\mathfrak{C}_{cor}|$ as we aim to establish a more precise
evaluation for the different cases that are possible. However,
in table \ref{tab:The-number-of} we report the number of names that
were found in the WoS data for each $|\mathfrak{C}_{cor}|$, from
which one can approximate the performance over the whole collection.
As the performance of our approach does not vary to any relevant extend
between using \emph{pairF1} or \emph{bCube}, we show only plots for
\emph{bCube}.

\subsection{Experiments}

In our experiments, we use the setup and the measures described above in combination with different parameters,
hyper-parameters and variants. Our model has the following variants:
\begin{enumerate}
\item \emph{within} / \emph{overall}: $\#(f)$ only within one name or over
all
\item \emph{pc\_on} / \emph{pc\_off}: using $p(C,\dot{C})$ or using $p(C|\dot{C})$
\item \emph{prob} / \emph{max}: sum-of-products or maximum-of-products
\end{enumerate}
As indicated earlier, for most variants, results show 
that only one of the two options was worth further investigation.
We choose the following setup based on first experiments on the training
data: (1) \emph{overall} and (2) \emph{pc\_off}. The difference between
(3) \emph{prob} and \emph{max} was not as clear. We decided that the
\emph{max} variant is worth further investigation but focused mostly
on the \emph{prob} variant as it could be implemented to run much
faster and first results also gave a better F1 score. Our model has
the following (hyper-) parameters:
\begin{enumerate}
\item Smoothing hyper-parameter $\epsilon$
\item Feature-type weights $\lambda$
\item Convergence parameters $\alpha$ and $\beta$
\end{enumerate}
So far, we have tried only one very small smoothing parameter $\epsilon=.0001$ to avoid division by zero.
We train the feature-type weights over the union of training portions for all
clustering sizes and approximate the results as shown in table \ref{tab:Feature-type-weights-considered}.
This table also shows all other feature-type weights examined.
Observing stopping size vs. correct size, we tuned the limit parameters in a 
manual grid-search on the training data and found that
a good choice is to set $\alpha=.0005$ for the \emph{max} variant
and $\beta=.000075$ for the \emph{prob} variant.
Our plots include a histogram of the clustering sizes $|\mathfrak{C}_{cor}|$
where the system clustering converged. All plots are given
for the training data as they have been part of the parameter tuning
process.

\subsection{Results}

\begin{table}
\begin{centering}
\tabcolsep=0.0pt{\footnotesize{}}%
\begin{tabular}{c|c|c|c|c|c|c|c|}
{\small{}$\begin{array}{r}
\mbox{\emph{term}}\\
\mbox{\emph{aff}}\\
\mbox{\emph{cat}}\\
\mbox{\emph{key}}\\
\mbox{\emph{co}}\\
\mbox{\emph{ref}}\\
\mbox{\emph{email}}\\
\mbox{\emph{year}}
\end{array}$} & {\small{}$\begin{array}{c}
.15\\
.2\\
.18\\
.03\\
.2\\
.12\\
.1\\
.02
\end{array}$} & {\small{}$\begin{array}{c}
.12\\
.03\\
.1\\
.2\\
.02\\
.15\\
.18\\
.2
\end{array}$} & {\small{}$\begin{array}{c}
.125\\
.125\\
.125\\
.125\\
.125\\
.125\\
.125\\
.125
\end{array}$} & {\small{}$\begin{array}{c}
1\\
0\\
0\\
0\\
0\\
0\\
0\\
0
\end{array}
\ddots
\begin{array}{c}
0\\
0\\
0\\
0\\
0\\
0\\
0\\
1
\end{array}$} & {\small{}$\begin{array}{c}
0\\
.143\\
.143\\
.143\\
.143\\
.143\\
.143\\
.143
\end{array}\ddots\begin{array}{c}
.143\\
.143\\
.143\\
.143\\
.143\\
.143\\
.143\\
0
\end{array}$} & {\small{}$\begin{array}{c}
0\\
0\\
0\\
0\\
.5\\
.5\\
0\\
0
\end{array}$} & {\small{}$\begin{array}{c}
.2\\
0\\
.2\\
.2\\
.2\\
0\\
0\\
.2
\end{array}$}\tabularnewline
 & \textbf{\emph{\footnotesize{}train}} & \textbf{\emph{\footnotesize{}opp.}} & \textbf{\emph{\footnotesize{}unif.}} & \textbf{\emph{\footnotesize{}select}} & \textbf{\emph{\footnotesize{}leaving-one-out}} & \textbf{\emph{\footnotesize{}author}} & \textbf{\emph{\footnotesize{}doc.}}\tabularnewline
\end{tabular}
\par\end{centering}{\footnotesize \par}
\caption{Feature-type weights considered in our experiments \label{tab:Feature-type-weights-considered}}
\end{table}

As mentioned above, we record a number of measurements during the clustering process
in order to understand the behavior of our method and the effect of different versions,
stopping limits and feature-type weightings. Table \ref{tab:test_results_bCube+pairF1} shows the final
results with tuned parameters on the test portion. In addition to that, we present our measurements in two types of plots. 
Referring to figure \ref{fig:trained_details} as an example, we briefly explain how to read the more comprehensive type: 
The y-axis displays the interval $[0,1]$, onto which precision, recall and
F1 are mapped. The x-axis gives the number of clusters in a clustering iteration. For a single block, 
clustering starts somewhere on the left of the plot with very low recall and maximal precision. 
As the process continues to merge clusters, recall increases and precision decreases. 
Over all blocks, we see the same development, but averaged for each problem size (standart deviation shown). 
In a perfect method, precision would remain constant until the correct number of clusters, shown as a solid vertical line, is reached. 
In figure \ref{fig:trained_details}, we see that for the \emph{max} variant, F1 peaks exactly at this point, while a bit later for the \emph{prob} variant.  
For both variants, the empirical stopping size (our method does not know the correct number of clusters) for $|\mathfrak{C}|=5$ 
peaks where F1 is maximal. The offset of this empirical distribution is tuned with the stopping 
parameters $\alpha$ and $\beta$. We can see in figure \ref{fig:trained_details_10} that there is still room for improvement of the 
stopping limit $l$, as for $|\mathfrak{C}|=10$ our method generally stops later than it should.

As preliminary tests on the training data clearly favour one combination
of variants and the stopping parameters are easily tuned for to these two options, 
the most interesting comparison is between different feature-type weightings.
As a first choice, trained weights from the classifier are used and
give satisfying results (see fig. \ref{fig:trained_results_prob}).
The \emph{max} variant performs worse in terms of F1, but precision
is higher (see fig. \ref{fig:trained_results_max}). Detailed plots
(fig. \ref{fig:trained_details}) of the clustering process suggest
that the \emph{prob} variant can sometimes gain relatively much recall
in early stages of the clustering, while the \emph{max} variant has
particularly regular behavior with smaller deviations from the mean.
It is also interesting to see that the \emph{max} variant is able
to gain the maximal F1 at the correct number of clusters, while the
\emph{prob} variant achieves higher values of F1 (due to better recall)
but peaks at a clustering size larger than the correct one. The average
maximum recall (\emph{'max rec.'}) at perfect precision (which is
independent of the stopping parameters) shown in figures \ref{fig:trained_results_prob}
and \ref{fig:trained_results_max} is much higher with the \emph{prob}
variant, which supports the notion that the \emph{prob} variant has
its strengths in a high recall. On the other hand, the average maximum
precision (\emph{'max prec.'}) at perfect recall is slightly higher
for the \emph{max} variant, suggesting that in general it can keep
precision higher until the end. We also show precision and F1 for
the baseline of putting all mentions into the same cluster (\emph{'base
prec.'} and \emph{'base f1'}). Obviously, recall is 1 here.

\begin{table*}[!t]
\begin{centering}
\tabcolsep=4.0pt\renewcommand{\arraystretch}{1.0}%
\begin{tabular}{|l|cccccccccc|ccccccc|cccccccccc|c|}
\multicolumn{1}{l|}{} & \textbf{1} & \multicolumn{1}{c}{\textbf{2}} & \multicolumn{1}{c}{\textbf{3}} & \multicolumn{1}{c}{\textbf{4}} & \multicolumn{1}{c}{\textbf{5}} & \multicolumn{1}{c}{\textbf{6}} & \multicolumn{1}{c}{\textbf{7}} & \multicolumn{1}{c}{\textbf{8}} & \multicolumn{1}{c}{\textbf{9}} & \textbf{10} & \textbf{$|\mathfrak{C}|$} &  &  &  &  &  & \textbf{$|\mathfrak{C}|$} & \textbf{1} & \textbf{2} & \textbf{3} & \textbf{4} & \textbf{5} & \textbf{6} & \textbf{7} & \textbf{8} & \textbf{9} & \textbf{10} & \multicolumn{1}{c}{}\tabularnewline
\hline 
\multirow{12}{*}{\begin{turn}{90}
\textbf{bCube}
\end{turn}} & \emph{\small{}100} & \emph{\small{}95} & \emph{\small{}95} & \emph{\small{}96} & \emph{\small{}97} & \emph{\small{}96} & \emph{\small{}94} & \emph{\small{}96} & \emph{\small{}95} & \emph{\small{}94} & \textbf{\small{}P} & \multirow{3}{*}{\begin{turn}{90}
\small{}trained
\end{turn}} & \multirow{6}{*}{\begin{turn}{90}
$\alpha=0$
\end{turn}} & \multirow{6}{*}{\begin{turn}{90}
\textbf{prob}
\end{turn}} & \multirow{6}{*}{\begin{turn}{90}
$\beta=.000075$
\end{turn}} & \multirow{3}{*}{\begin{turn}{90}
\small{}trained
\end{turn}} & \textbf{\small{}P} & \emph{\small{}100} & \emph{\small{}95} & \emph{\small{}95} & \emph{\small{}96} & \emph{\small{}96} & \emph{\small{}96} & \emph{\small{}94} & \emph{\small{}95} & \emph{\small{}94} & \emph{\small{}93} & \multirow{12}{*}{\begin{turn}{90}
\textbf{pairF1}
\end{turn}}\tabularnewline
 & \textbf{\emph{98}} & \textbf{\emph{93}} & \textbf{\emph{93}} & \textbf{\emph{92}} & \textbf{\emph{91}} & \textbf{\emph{90}} & \textbf{\emph{91}} & \textbf{\emph{91}} & \textbf{\emph{90}} & \textbf{\emph{89}} & \textbf{F} &  &  &  &  &  & \textbf{F} & \textbf{\emph{99}} & \textbf{\emph{94}} & \textbf{\emph{93}} & \textbf{\emph{92}} & \textbf{\emph{92}} & \textbf{\emph{91}} & \textbf{\emph{92}} & \textbf{\emph{92}} & \textbf{\emph{91}} & \emph{90} & \tabularnewline
 & \emph{\small{}97} & \emph{\small{}91} & \emph{\small{}90} & \emph{\small{}88} & \emph{\small{}86} & \emph{\small{}85} & \emph{\small{}87} & \emph{\small{}87} & \emph{\small{}86} & \emph{\small{}84} & \textbf{\small{}R} &  &  &  &  &  & \textbf{\small{}R} & \emph{\small{}97} & \emph{\small{}92} & \emph{\small{}92} & \emph{\small{}89} & \emph{\small{}88} & \emph{\small{}87} & \emph{\small{}90} & \emph{\small{}89} & \emph{\small{}88} & \emph{\small{}87} & \tabularnewline
\cline{2-13} \cline{17-28} 
 & \emph{\small{}100} & \emph{\small{}95} & \emph{\small{}95} & \emph{\small{}97} & \emph{\small{}97} & \emph{\small{}97} & \emph{\small{}95} & \emph{\small{}96} & \emph{\small{}96} & \emph{\small{}95} & \textbf{\small{}P} & \multirow{3}{*}{\begin{turn}{90}
\small{}uniform
\end{turn}} &  &  &  & \multirow{3}{*}{\begin{turn}{90}
\small{}uniform
\end{turn}} & \textbf{\small{}P} & \emph{\small{}100} & \emph{\small{}95} & \emph{\small{}95} & \emph{\small{}96} & \emph{\small{}97} & \emph{\small{}96} & \emph{\small{}94} & \emph{\small{}96} & \emph{\small{}94} & \emph{\small{}94} & \tabularnewline
 & \textbf{\emph{98}} & \textbf{\emph{93}} & \textbf{\emph{93}} & \textbf{\emph{92}} & \emph{90} & \textbf{\emph{90}} & \emph{90} & \emph{90} & \emph{89} & \textbf{\emph{89}} & \textbf{F} &  &  &  &  &  & \textbf{F} & \textbf{\emph{99}} & \textbf{\emph{94}} & \textbf{\emph{93}} & \textbf{\emph{92}} & \emph{91} & \emph{91} & \emph{91} & \emph{91} & \emph{90} & \textbf{\emph{91}} & \tabularnewline
 & \emph{\small{}97} & \emph{\small{}91} & \emph{\small{}90} & \emph{\small{}87} & \emph{\small{}84} & \emph{\small{}84} & \emph{\small{}85} & \emph{\small{}85} & \emph{\small{}83} & \emph{\small{}84} & \textbf{\small{}R} &  &  &  &  &  & \textbf{\small{}R} & \emph{\small{}97} & \emph{\small{}92} & \emph{\small{}91} & \emph{\small{}89} & \emph{\small{}86} & \emph{\small{}85} & \emph{\small{}88} & \emph{\small{}87} & \emph{\small{}85} & \emph{\small{}87} & \tabularnewline
\cline{2-28} 
 & \emph{\small{}100} & \emph{\small{}96} & \emph{\small{}96} & \emph{\small{}97} & \emph{\small{}96} & \emph{\small{}95} & \emph{\small{}92} & \emph{\small{}92} & \emph{\small{}91} & \emph{\small{}88} & \textbf{\small{}P} & \multirow{3}{*}{\begin{turn}{90}
\small{}trained
\end{turn}} & \multirow{6}{*}{\begin{turn}{90}
$\alpha=.0005$
\end{turn}} & \multirow{6}{*}{\begin{turn}{90}
\textbf{max}
\end{turn}} & \multirow{6}{*}{\begin{turn}{90}
$\beta=0$
\end{turn}} & \multirow{3}{*}{\begin{turn}{90}
\small{}trained
\end{turn}} & \textbf{\small{}P} & \emph{\small{}100} & \emph{\small{}96} & \emph{\small{}96} & \emph{\small{}97} & \emph{\small{}96} & \emph{\small{}95} & \emph{\small{}92} & \emph{\small{}90} & \emph{\small{}89} & \emph{\small{}87} & \tabularnewline
 & \emph{96} & \emph{92} & \emph{91} & \emph{91} & \emph{90} & \textbf{\emph{90}} & \emph{89} & \emph{89} & \emph{89} & \emph{88} & \textbf{F} &  &  &  &  &  & \textbf{F} & \emph{97} & \emph{92} & \emph{92} & \emph{92} & \emph{90} & \emph{90} & \emph{89} & \emph{88} & \emph{88} & \emph{88} & \tabularnewline
 & \emph{\small{}93} & \emph{\small{}88} & \emph{\small{}87} & \emph{\small{}86} & \emph{\small{}84} & \emph{\small{}85} & \emph{\small{}87} & \emph{\small{}86} & \emph{\small{}87} & \emph{\small{}88} & \textbf{\small{}R} &  &  &  &  &  & \textbf{\small{}R} & \emph{\small{}94} & \emph{\small{}89} & \emph{\small{}88} & \emph{\small{}87} & \emph{\small{}84} & \emph{\small{}85} & \emph{\small{}87} & \emph{\small{}86} & \emph{\small{}86} & \emph{\small{}89} & \tabularnewline
\cline{2-13} \cline{17-28} 
 & \emph{\small{}100} & \emph{\small{}96} & \emph{\small{}96} & \emph{\small{}97} & \emph{\small{}96} & \emph{\small{}95} & \emph{\small{}92} & \emph{\small{}92} & \emph{\small{}91} & \emph{\small{}88} & \textbf{\small{}P} & \multirow{3}{*}{\begin{turn}{90}
\small{}uniform
\end{turn}} &  &  &  & \multirow{3}{*}{\begin{turn}{90}
\small{}uniform
\end{turn}} & \textbf{\small{}P} & \emph{\small{}100} & \emph{\small{}96} & \emph{\small{}96} & \emph{\small{}97} & \emph{\small{}96} & \emph{\small{}94} & \emph{\small{}91} & \emph{\small{}91} & \emph{\small{}89} & \emph{\small{}87} & \tabularnewline
 & \emph{96} & \emph{92} & \emph{91} & \emph{91} & \emph{89} & \textbf{\emph{90}} & \emph{89} & \emph{89} & \emph{88} & \emph{88} & \textbf{F} &  &  &  &  &  & \textbf{F} & \emph{97} & \emph{92} & \emph{92} & \emph{91} & \emph{89} & \emph{89} & \emph{89} & \emph{85} & \emph{87} & \emph{88} & \tabularnewline
 & \emph{\small{}92} & \emph{\small{}88} & \emph{\small{}87} & \emph{\small{}86} & \emph{\small{}83} & \emph{\small{}85} & \emph{\small{}86} & \emph{\small{}86} & \emph{\small{}85} & \emph{\small{}88} & \textbf{\small{}R} &  &  &  &  &  & \textbf{\small{}R} & \emph{\small{}93} & \emph{\small{}89} & \emph{\small{}87} & \emph{\small{}86} & \emph{\small{}83} & \emph{\small{}85} & \emph{\small{}87} & \emph{\small{}88} & \emph{\small{}84} & \emph{\small{}89} & \tabularnewline
\hline 
\end{tabular}
\par\end{centering}
\caption{Results of the tuned method on the test portion, using bCube and pairF1
measure \label{tab:test_results_bCube+pairF1}}
\end{table*}

The trained feature-type weighting is contrasted with a uniform weighting
of feature-types. Results are almost identical as can be seen in figure
\ref{fig:uniform}, suggesting that equal weighting of all feature-types
is a good choice. However, this does not mean that feature-type weighting
has no influence on the performance. In order to investigate the influence
of feature-type weighting, we test an 'opposed' weighting where the
feature-type with the previously smallest weight is assigned the previously
biggest weight, the feature-type with the previously biggest weight
is assigned the previously smallest weight, and so on (see table \ref{tab:Feature-type-weights-considered}).
Results are clearly worse (see fig. \ref{fig:opposed}), which shows
that feature-type weighting is not completely irrelevant (in the sense
that there are also counterproductive weightings). Furthermore, we
use feature-type weighting to investigate the effect of single feature-types
by on the one hand leaving them out (setting their weight to zero)
and on the other using them exclusively (setting all others to zero). 
This analysis shows that most feature-types can be dropped (if they
are the only one to be dropped) with the exception of co-authors (see fig. \ref{fig:leave-out-coauthors}) and
ref-authors. It is therefore interesting
to see the results if only these two feature-types are used (see fig.
\ref{fig:authors}). Performance is solid. The most important coreference indicators are co-author names, 
which might even be used alone (see fig. \ref{fig:select_coauthors}) . In a last experiment, we also test a
weighting that selects only basic features of the document $d(x)$
(no features associated with the author mention $x$ itself and no
referenced authors). Results (fig. \ref{fig:docfeats}) are
acceptable: F1 is not as good as before, but precision is high.

\subsection{Discussion}

A direct comparison of our experimental results to
those obtained by other researchers is not possible. In fact such a
comparison would almost certainly not be fair as evaluation results
depend heavily on so many different factors that reproducibility is
close to impossible under normal circumstances:
\begin{enumerate}
\item \emph{Dataset}: size, distribution of authors, version of data set, domain, availability of features, completeness of author name specification, hand-selected?, quality and amount of gold-annotation
\item \emph{Blocking scheme}: average size of blocks, unassigned names, overlapping blocks?
\item \emph{Evaluation measure}: general choice of measure, micro/macro average?, recall only inside block?, pair- or element-wise comparison?, counting pairs of equal mentions?, evaluated for different problem sizes?
\end{enumerate}
It is therefore more than desirable to have a benchmark dataset with
a framework that distinguishes clearly between \emph{data}, \emph{annotation}, \emph{blocking},
\emph{disambiguation} and \emph{evaluation}. 
As a such benchmark is not available at a realistic scale like the Web of Science, we promote
the paradigm to evaluate different problem sizes (number of clusters that
need to be found) separately. This makes our results relatively robust
against changes to many (not all) of the above mentioned factors.
Unfortunately, this kind of evaluation can not be referenced from other
publications that have worked on the Web of Science. Therefore, we base
the assessment of the AD method proposed in this work on the fact that
we were able to achieve results of more than 90\% F1 for problem sizes of at
least ten authors, which we find promising. Closely related work by Levin et al. \cite{levin2012citation}
or Gurney et al. \cite{gurney2012author} does not report much higher
values even though they do not separate the problems sizes so rigorously
-- which generally improves the results. 
However, we must note that Levin et al. cluster blocks of surname and first initial, 
while we use \emph{all} initials. Therefore, they work with much larger problem sizes, 
which is computationally challenging and leads to more realistic 
estimation of recall, even if measured only within blocks. We hope that reporting 
results for individual problem sizes minimizes these distortions in the comparison.

\section{Conclusions}

In the following, we briefly summarize the main findings of our research:
The best variant is using $p(C|\dot{C})$ instead of $p(C,\dot{C})$
with $\#(f)$ over the whole collection. Using maximum-of-products
instead of sum-of-products constitutes a serious alternative. There
are some hints that it might be even more precise (\emph{'max prec.'}
is slightly higher). Certainly, it runs significantly slower in our
implementation, which is why we have not yet fully investigated potential
gains of fine-tuning this variant. When tuning an appropriate stopping
parameter, our method can deliver state-of-the-art results although
being conceptionally simple. Even though feature-type weights learned in
the classification scenario are quite heterogeneous, when applied in
the clustering application, they do not perform better than uniformly
distributed weights. We view this as a benefit of our model, as its
score works well independent of any training. We hypothesize that
the probabilities filter out unspecific features, thereby implicitly
controlling feature weighting without any discriminative training.
The stopping criterion needs to be tuned on some training set, but
it is only a single variable per variant that needs to be fitted. This
quality limit does remarkably well at finding an appropriate number
of clusters to converge. Leaving out one feature-type at a time in
the clustering score, we find that co-author names are the most
important features. Apart from this, our method's performance
is not dependent on the presence of specific feature-types. For example,
a weighting where only the co-author names and the referenced author
names are used performs well. Recording the results for each
system clustering size $|\mathfrak{C}_{sys}|$ separately allows to
precisely monitor the behavior of the clustering process. The plots
created in this work also allow tuning precision vs. recall. This
might be particularly interesting for digital libraries, as they might
prefer precision over recall. Separate evaluation for each
correct clustering size $|\mathfrak{C}_{cor}|$ shows how high
the baseline of putting all mentions in a single cluster is for the
frequent cases of $|\mathfrak{C}_{cor}|=1$ or $|\mathfrak{C}_{cor}|=2$.
We conclude that, with larger problems not separated, an approach
could easily be considered satisfying although only approximating
this primitive baseline.

%% file: AD_algorithm.tex
\begin{algorithm}[t]
\SetAlgoNoEnd
\SetKwInOut{Input}{Input}
\Input{$name$ with $X$ and $\mathcal{C}$ as well as $N,\#(f),\lambda,\epsilon,l$}
\While{$|\mathcal{C}|>1$}{
	$score   \leftarrow NULL$ \;
	\ForEach{$(C,\dot{C})\in\mathcal{C}\times\mathcal{C}$}{
		$score(C,\dot{C})\leftarrow\sum_{ftype}\lambda_{ftype}\cdot p_{ftype}(C|\dot{C})$ \;
	}
	$merges \leftarrow \{\}$ \;
	\ForEach{$(C,\dot{C})\in\mathcal{C}\times\mathcal{C}$}{
		\If{$\neg\exists \ddot{C}\in\mathcal{C}:score(C,\ddot{C})>score(C,\dot{C})$ \textbf{and} $\neg\exists \dddot{C}\in \mathcal{C}:score(\dddot{C},\dot{C})>score(C,\dot{C})$ \textbf{and} $score(C,\dot{C})>l$}{
			$merges \leftarrow merges \cup (C,\dot{C})$ \;
		}
	}
	\lIf{$|merges|$ = 0}{
		break
	}
	\Else{
		\ForEach{$(C,\dot{C})\in merges$}{
			$merge((C,\dot{C}))$ \;
		}
	}
}
\caption{Our agglomerative clustering (no evaluation)\label{alg:1}}
\end{algorithm}

%% file: AD_appendix.tex
\begin{figure}
\begin{centering}
\includegraphics[width=.91\columnwidth]{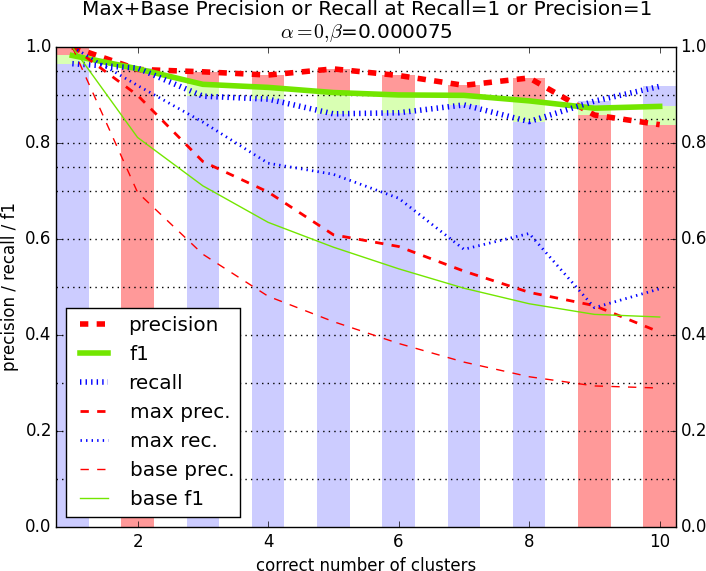}
\par\end{centering}
\caption{Results for trained weights $\lambda$;using \emph{prob} variant\label{fig:trained_results_prob}}
\end{figure}

\begin{figure}
\begin{centering}
\includegraphics[width=.91\columnwidth]{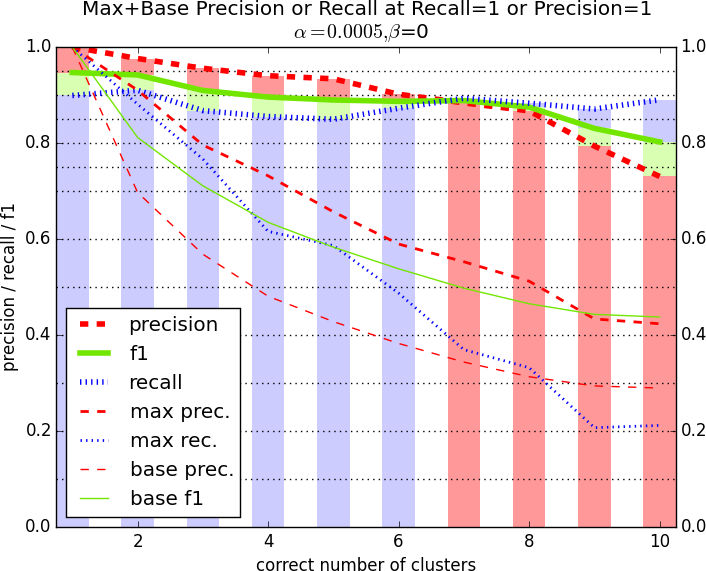}
\par\end{centering}
\caption{Results for trained weights $\lambda$; using \emph{max }variant\label{fig:trained_results_max}}
\end{figure}

\begin{figure}
\begin{centering}
\includegraphics[width=.91\columnwidth]{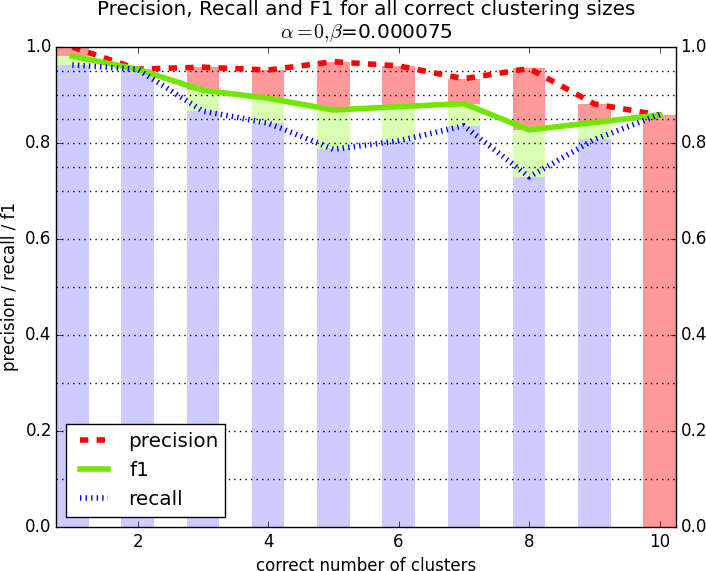}
\par\end{centering}
\caption{Results for opposed weights (compare figure \ref{fig:trained_results_prob})\label{fig:opposed}}
\end{figure}

\begin{figure*}
\begin{centering}
\includegraphics[width=.91\columnwidth]{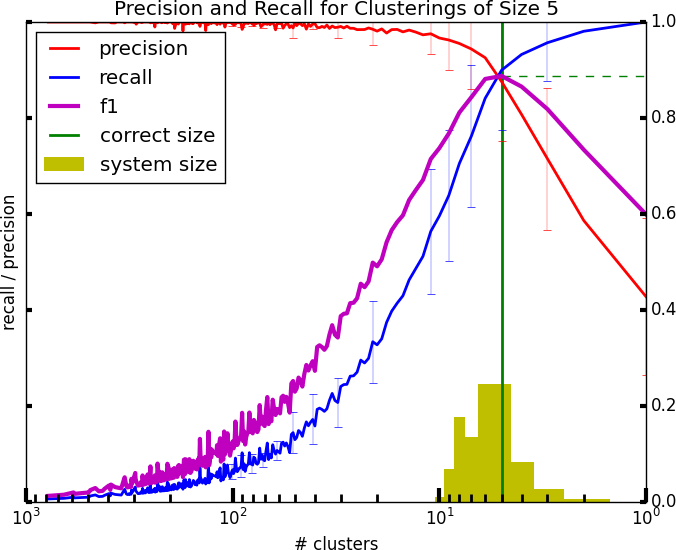}\hspace{25pt}\includegraphics[width=.91\columnwidth]{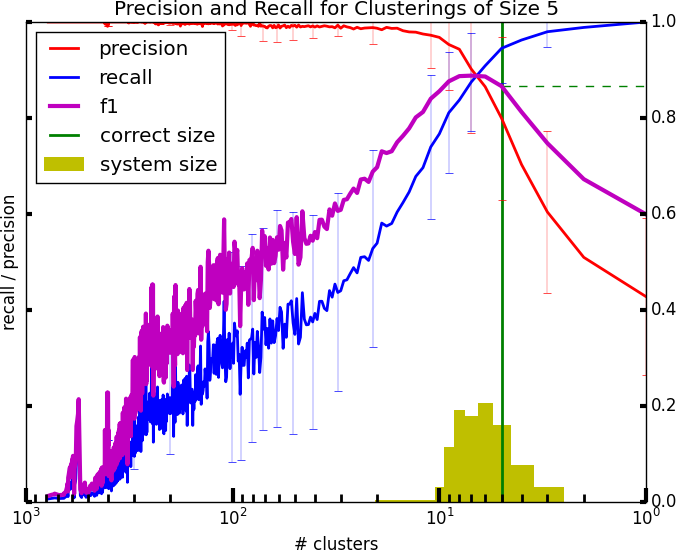}
\par\end{centering}
\caption{The clustering process visualised for trained weights $\lambda$; comparing \emph{max} and \emph{prob} variant; $|\mathfrak{C}|=5$\label{fig:trained_details}}
\end{figure*}

\begin{figure*}
\begin{centering}
\includegraphics[width=.91\columnwidth]{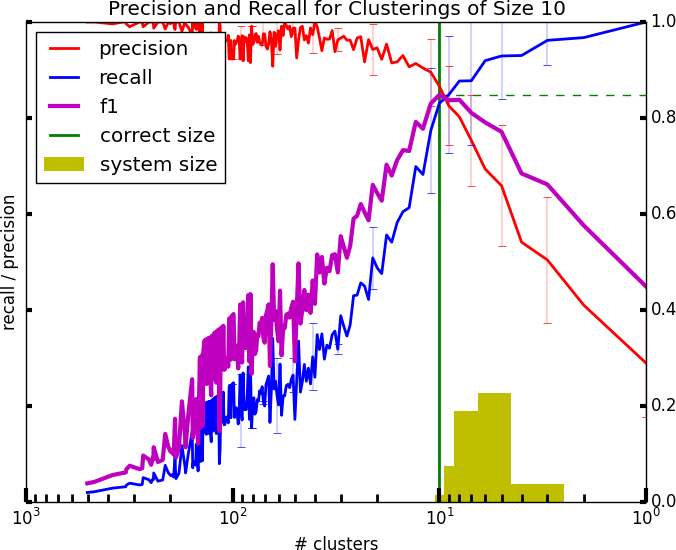}\hspace{25pt}\includegraphics[width=.91\columnwidth]{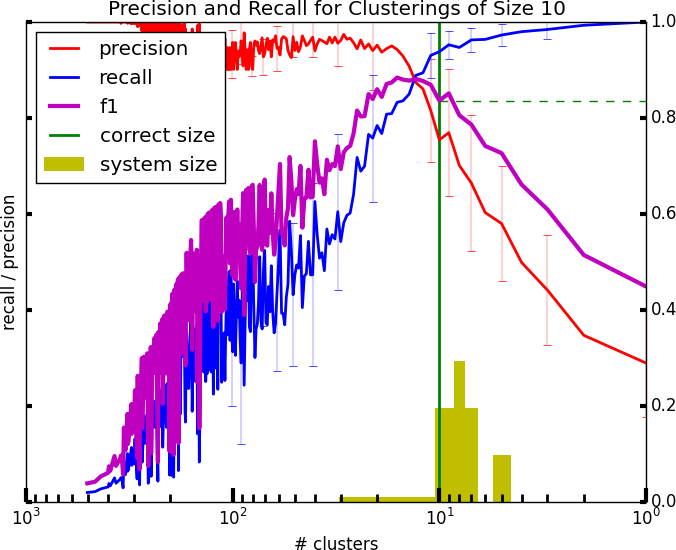}
\par\end{centering}
\caption{The clustering process visualised for trained weights $\lambda$; comparing \emph{max} and \emph{prob} variant; $|\mathfrak{C}|=10$\label{fig:trained_details_10}}
\end{figure*}

\begin{figure*}
\begin{centering}
\includegraphics[width=.91\columnwidth]{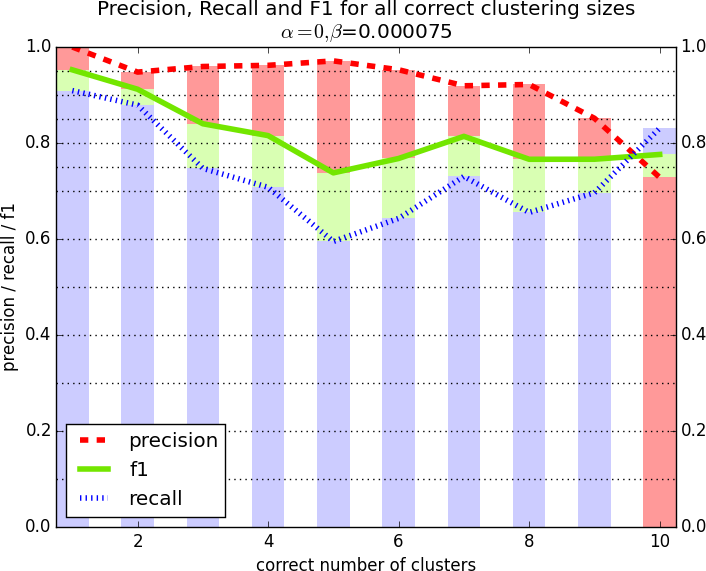}\hspace{25pt}\includegraphics[width=.91\columnwidth]{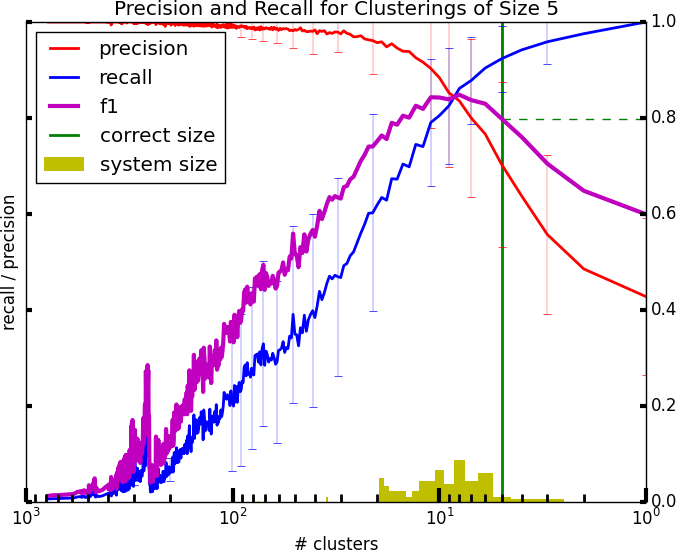}
\par\end{centering}
\caption{Selecting only terms as features; results and details of the clustering
process for $|\mathfrak{C}|=5$ \label{fig:select_terms}}
\end{figure*}

\begin{figure*}
\begin{centering}
\includegraphics[width=.91\columnwidth]{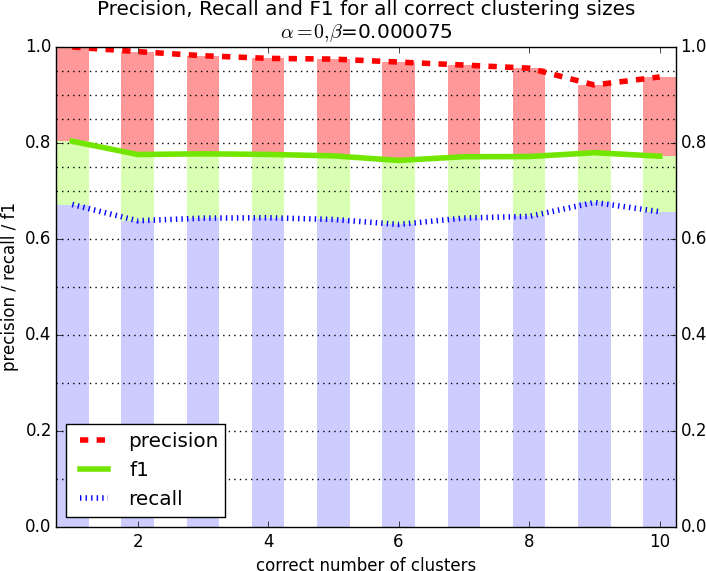}\hspace{25pt}\includegraphics[width=.91\columnwidth]{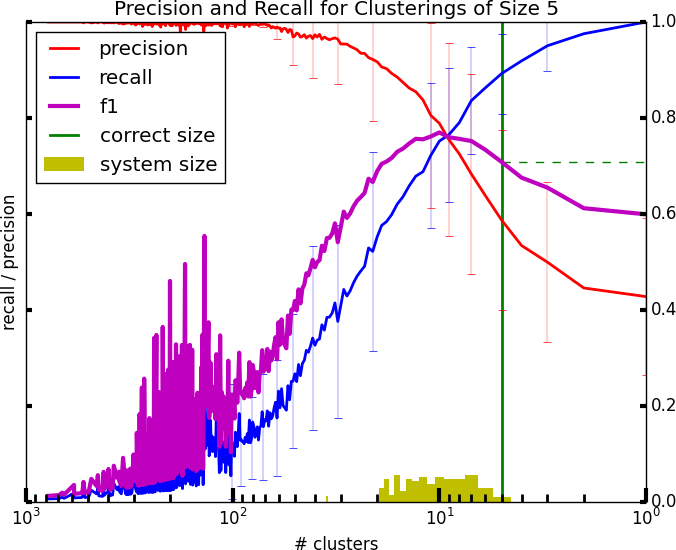}
\par\end{centering}
\caption{Selecting only co-authors as features; results and details for $|\mathfrak{C}|=5$
\label{fig:select_coauthors}}
\end{figure*}

\begin{figure}
\begin{centering}
\includegraphics[width=.91\columnwidth]{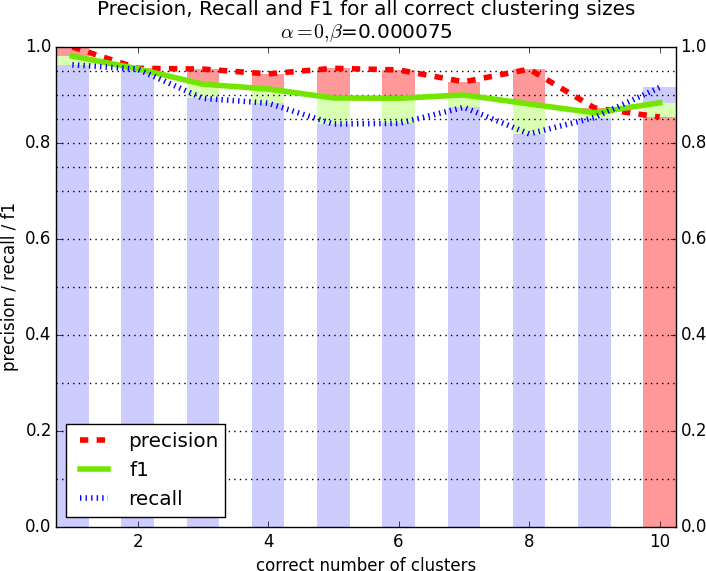}
\par\end{centering}
\caption{Results for uniform weights (compare figure \ref{fig:trained_results_prob})\label{fig:uniform}}
\end{figure}

\begin{figure}
\begin{centering}
\includegraphics[width=.91\columnwidth]{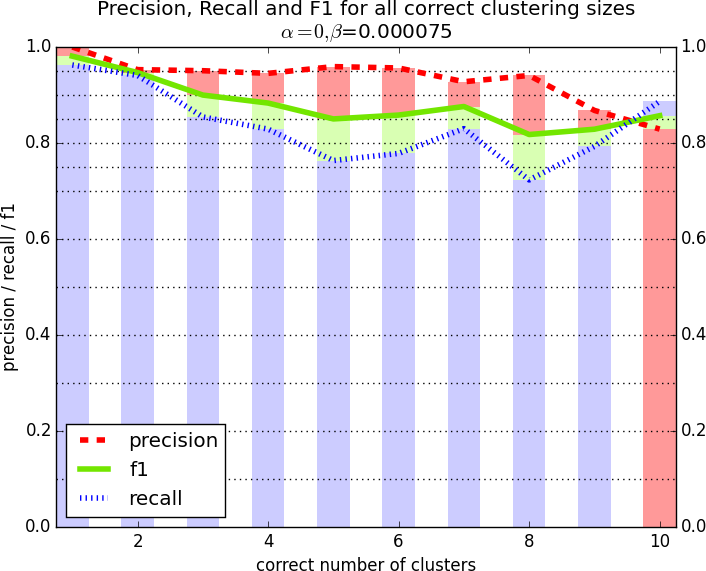}
\par\end{centering}
\caption{Leaving out co-authors (compare figure \ref{fig:trained_results_prob})\label{fig:leave-out-coauthors}}
\end{figure}

\begin{figure}
\begin{centering}
\includegraphics[width=.91\columnwidth]{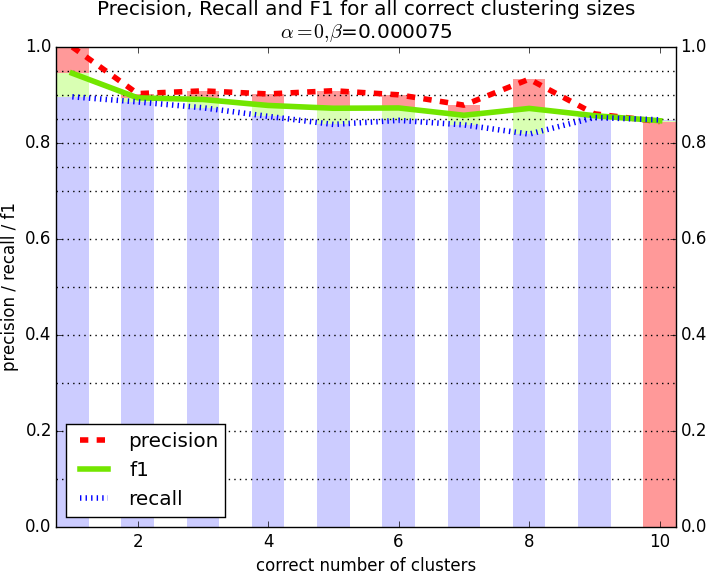}
\par\end{centering}
\caption{Only co- and referenced authors (compare fig. \ref{fig:trained_results_prob})\label{fig:authors}}
\end{figure}

\begin{figure}
\begin{centering}
\includegraphics[width=.91\columnwidth]{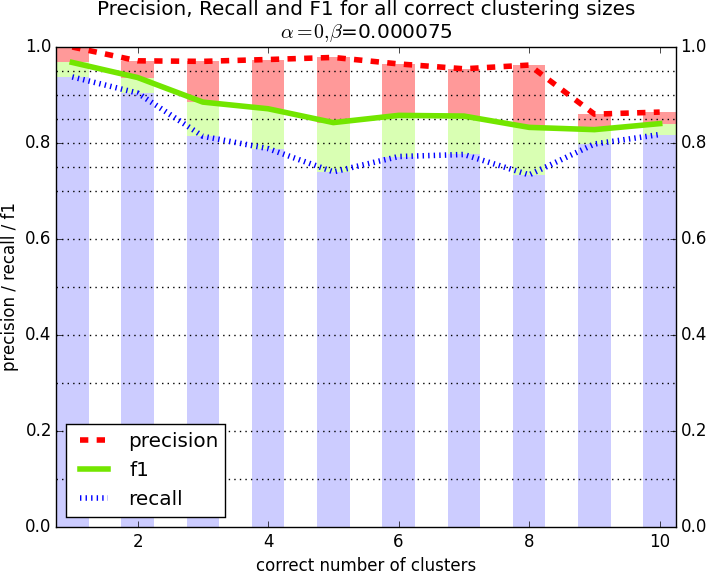}
\par\end{centering}
\caption{No author-specific features (compare figure \ref{fig:trained_results_prob})\label{fig:docfeats}}
\end{figure}